# Simulating Organogenesis in COMSOL: Cell-based Signaling Models

Jannik Vollmer[1,2], Denis Menshykau[1,2], Dagmar Iber[*,1,2]
[1]D-BSSE, ETH Zurich, Switzerland; [2]SIB, Switzerland
*Corresponding author: D-BSSE, ETH Zurich, Mattenstrasse 26, 4058, Switzerland, dagmar.iber@bsse.ethz.ch

**Abstract:** Most models of biological pattern formation are simulated on continuous domains even though cells are discrete objects that provide internal boundaries to the diffusion of regulatory components. In our previous papers on simulating organogenesis in COMSOL (Germann et al COMSOL Conf Procedings 2011; Menshykau and Iber, COMSOL Conf Proceedings 2012) we discussed methods to efficiently solve signaling models on static and growing continuous domains. Here we discuss COMSOL-based methods to study spatio-temporal signaling models at cellular resolution with subcellular compartments, i.e. cell membrane, cytoplasm, and nucleus.

**Keywords:** cell compartments, reaction-diffusion, moving and deforming domain, COMSOL.

## 1. Introduction

Organogenesis is a tightly regulated process that has been studied for decades. While many regulatory genes have been identified, the underlying regulatory networks are too complex to be grasped by verbal models alone and the control mechanisms are therefore only poorly understood. Computational models can help to fill this gap. They may help to distinguish between various proposed mechanisms, to generate new hypotheses and to design experiments supporting or rejecting these hypotheses [1].

During organogenesis tissue layers organize and differentiate into functionally organized units. Most developmental processes are remarkably stereotyped, such that patterns are identical in developing embryos of the same genetic background, i.e. the same lung branching patterns are observed in littermate mice with the same genetic background, except for branching errors [2, 3]. Deterministic models are therefore an appropriate method to describe these processes and have been frequently used. We have previously developed and solved such models in COMSOL Multiphysics [4, 5]. These simulations allowed us to correctly predict novel genetic regulatory interactions [6], and to propose new regulatory mechanisms. In particular we proposed that spontaneous, yet deterministic symmetry breaks in a pattern may result from a receptor-ligand based Turing mechanism. We applied the mechanism to both digit patterning in the limb, and branch point selection during branching morphogenesis [7, 8,9].

In all of these simulations we approximated tissue as a continuous domain. However, in particular in case of the receptor-ligand based Turing mechanism a cell resolution may be important as receptors are restricted to single cells. The cell boundaries, and in some cases even the subcellular compartmentalization, may have a crucial effect on the signaling networks. This makes it necessary to develop methods to appropriately simulate models, which describe these processes on the cellular level. Here we discuss COMSOL-based implementations to study spatio-temporal signaling models at cellular resolution.

In order to implement cell-based signaling models in COMSOL we need to be able to restrain model species and reactions to specific compartments, while coupling them to other reactions. COMSOL offers several different ways to implement cell-based models. Here we summarize and evaluate the accuracy and performance of these different approaches. Going forward, the described tools can be used to expand the continuous models to small tissue models consisting of multiple cells.

## 2. Methods
### 2.1 General modeling approach

Our models for morphogenesis are formulated as coupled systems of reaction-diffusion equations of the form:

$$\frac{\partial X_i}{\partial t} = D_i \nabla^2 X_i + R_i$$

where $D_i$ is the diffusion constant of component $i$

and $\nabla$ is the Nabla operator. $R_i$ denotes the reactions, which couple the equations for the different species $X_i$. A wide range of reaction laws can be used; typical reactions include $R_X = -\delta \cdot X$ for the decay of component X and

$$R_X = -k^+ \cdot m \cdot X^m \cdot Y^n + k^- \cdot m \cdot X_m Y_n$$
$$R_Y = -k^+ \cdot n \cdot X^m \cdot Y^n + k^- \cdot n \cdot X_m Y_n$$
$$R_{X_m Y_n} = k^+ \cdot X^m \cdot Y^n - k^- \cdot X_m Y_n$$

for the formation of a complex $X_m Y_n$ made of $m$ $X$ and $n$ $Y$ molecules. The reaction terms can contain also other non-linear functions like enzymatic activation $\sigma$ and inhibition $\bar{\sigma} = 1 - \sigma$, where $\sigma$ is modelled analogous to Hill kinetics (Michaelis-Menten for n=1):

$$\sigma = X^n / (X^n + K^n).$$

The threshold $K$ is the concentration at which the activation reaches half its strength and the exponent $n$ depends on the cooperativity of the regulating interactions. For example $R_X = \rho \cdot \sigma(Y)$ describes a production term for a protein $X$ induced by another protein $Y$.

As some models within this work do not show any spatial gradients, an ordinary differential equation-based approach would have normally been more suitable for those. However, in order to be able to compare the accuracy and performance of all models, these models were also implemented as reaction-diffusion models.

## 2.2 Computational Details

All models discussed here were implemented in COMSOL 4.3b using the *Coefficient Form PDE* or *Coefficient Form Boundary PDE* interfaces. If not stated otherwise, models were solved using the *direct* solver *MUMPS* with its default solver options and a *normal* mesh size. Variables were always fully coupled.

## 3. Coupling of different compartments

Similar to the implementation of different tissue layers in COMSOL [4], models with cellular or subcellular resolution can, in principle, be implemented in two different ways: as set of global PDEs with different parameters on the different domains or by separated PDEs, which are subsequently coupled. We compared the different implementations and observed that the first approach is not suitable for our models.

To compare the two different implementations, we chose a simple 2-dimensional model, in which a freely diffusible ligand $L$ binds to a cell-surface receptor $R$. Assuming the law of mass action, the concentration of the complex $C$ on the surface of the cell can be described by:

$$\frac{\partial C}{\partial t} = D_C \nabla^2 C + k_{on} \cdot R \cdot L$$

where $k_{on}$ is a parameter describing the speed of the binding, and $R$ and $L$ are the concentrations of the these species.

The time derivates of species $L$ and $R$ can be described accordingly, but their concentration decreases with the increasing formation of the complex.

The geometry in this model consists of two squares with side length $s_1 = 1$ and $s_2 = 0.5$,

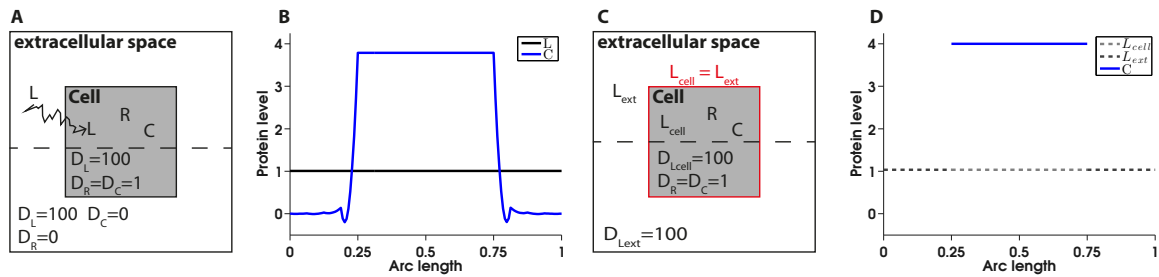

**Figure 1. Coupling of compartments.** A) Schematic representation of the 2D domain showing the localization of the species in the model and the values of the diffusion coefficients on the two domains. Ligand L is freely diffusible. B) Protein concentrations along the dashed line shown in (A). Artifacts in the concentration of complex *C* can be observed at the boundary of the two domains. C) Same geometry as in (A). However, Ligand *L* is split into two species which are present on the extracellular space and the cellular domain, respectively, and subsequently coupled. D) Using coupled PDEs restricts the diffusion of the complex *C* to the surface of the cell and does not show any artifacts.

respectively (Fig. 1A). The inner compartment describes the surface of the cell, while the surrounding compartment describes the extracellular space.

### 3.1 Implementation using a global set of PDEs

As a first approach, we define the PDEs globally for both domains with no flux boundary conditions (BCs) on the outermost boundaries. We assume that the diffusion of the receptor $R$ is restricted to the cell, while the ligand $L$ is freely diffusible (Fig. 1A). The parameters are chosen accordingly. Thus, on the inner compartment, corresponding to the cell, parameters are defined as $D_L=100$, $D_R=D_C=1$ and $k_{on}=0.1$, while $D_R=D_{RL}=k_{on}=0$ is assumed for the extracellular space.

$\int_E L\,dA = 2$ and $\int_C R\,dA = 1$ were chosen as initial conditions (ICs) for $L$ and $R$ respectively, with $E$ and $C$ describing the area of the extracellular space and the cell, respectively. Thus, at time point $t=0$, receptor and ligand are present only on the cell and only in the extracellular space, respectively.

When we solve this system in COMSOL Multiphysics with the options described in 2.2, we observe artifacts near to the border of the cell (Fig. 1B). Most importantly, this results in negative concentrations of the complex $C$.

### 3.2 Implementation using a set of coupled PDEs

As a set of global PDEs results in artifacts near the borders of the subdomains, we now define the PDEs separately for the two different types of domains. The PDEs are then subsequently coupled by imposing the boundary condition $L_{ext} = L_{cell}$ at the borders between cell and extracellular space such that these two species have always the same concentration at these boundaries (Fig. 1C). While using the same solver and ICs as in 3.1, the concentration profile no longer shows any artifacts (Fig. 1D). The species $R$ as well as the complex $C$ are restricted to the surface of the cell.

## 4. Surface Reactions

While the previous example allows to model complex formation on a cell, it does not allow to spatially model species downstream of this complex. In order to model such subcellular reactions downstream of ligand-receptor interactions, we need to introduce additional compartments.

Biologically, the cell membrane is formed by a lipid bilayer and is approximately 5-10 nm thick. It can therefore either be modeled as a thin compartment, equivalently to the previous example, or due to its small thickness, as a simple boundary of $n$-1 dimension, where $n$ is the dimension of the complete cell, i.e. a 1D boundary in our 2D example. To compare the different implementations, we considered three different models: In the first, the membrane was modeled explicitly as a separate compartment, which is coupled to the extracellular space using a constraint as described above (Fig. 2A). In the second and third, the membrane was defined as the 1D boundary between cytoplasm and extracellular space and implemented using the *coefficient form boundary PDE* interface (Fig. 2C,E). The models differed in the BCs implemented.

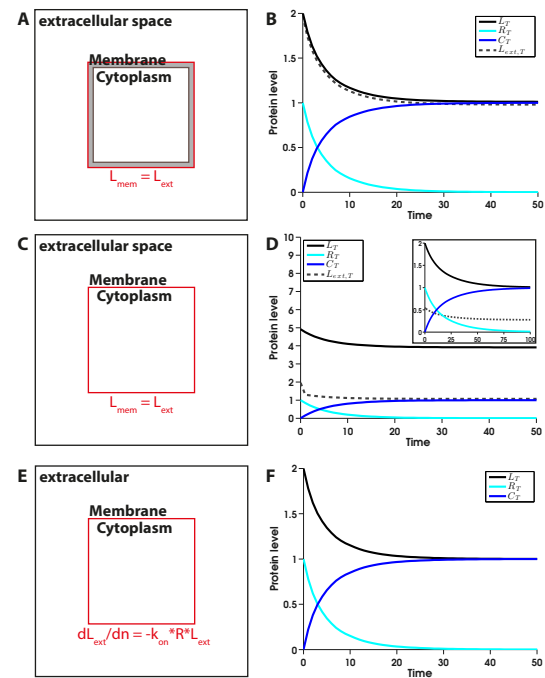

**Figure 2. Implementation of surface reactions.** A,B) Model geometry and time course for implementation of membrane as a separate compartment. C-E) Membrane implemented as boundary. The models differ in the coupling of the PDEs. PDEs are either coupled using concentration BCs as in A) (C,D) or via a flux BC (E,F). Coupling using constraints results in an altered time course and steady state (D), also if ICs are changed (D,inset), while use of flux BC results in the same time course as in A). All protein levels are given as sum of all subspecies integrated over their respective compartments.

The same BC as in the first model (constraint of concentrations) was used in the second model. In the third model, however, a flux BC was used. This flux BC defined that ligand $L$ was lost at the boundary between extracellular space and membrane following:

$$\frac{\partial L_{ext}}{\partial n} = -k_{on} \cdot R \cdot L_{ext}$$

where $n$ is a vector normal to the boundary, $L_{ext}$ describes the concentration of ligand directly at the boundary and $R_{Mem}$ the concentration of the receptor on the boundary. Receptor-ligand complex formation on the boundary was implanted equivalently using the **coeffient form boundary PDE** interface as:

$$\frac{\partial C}{\partial t} = k_{on} \cdot R \cdot L_{ext}$$

The initial conditions were chosen as previously such that only receptor was present on the cell membrane and only ligand in the extracellular space at time point $t$=0.

While the first and third implementation resulted in the same time courses for the concentration of all species (Fig.2 B,F), the second implementation showed artifacts. The concentration of ligand L was much too high, even higher than its initial concentration, when it was integrated over all domains (Fig. 2D).

A major influence can be observed for the IC in the case of the second model. If the IC for the ligand $L$ is chosen such that it is also present on the membrane at $t$=0, the time course for the integrated concentration over all domains, shows the expected result (Fig. 2D inset). However, the concentration in the domain describing the extracellular space is much too low and the model needs to be simulated much longer to reach its steady state. Thus, this implementation cannot be used for coupling of two compartments of different dimensions.

While the accuracy of the results was similar for the first and third implementation, a striking difference was observed for the time needed to compute the two models with the first model taking around 2 fold the time of the third (8 vs. 4 seconds) as the mesh needs to be much finer around the small layer of the membrane to describe the geometry accurately. Thus, the models differed in their degree of freedom (4864 vs. 1184 for normal mesh size and physics-controlled mesh). The same ratio in computational time was observed when the mesh size was made much finer (61 vs. 31 seconds and 44768 and 39676 degrees of freedom for extremely fine mesh).

## 4. A 3D cell-based signaling model

In order to test the performance of the two alternatives of modeling a membrane, i.e. as compartment with constraints, or as boundary with flux BC, we built a more complex 3D model.

In this model, as previously, ligand binds irreversibly to a receptor on a membrane. The resulting complex C then activates an intracellular species $X$. The activated form $X_p$, which rapidly becomes deactivated to its non-active form $X$, can then diffuse into the nucleus where it enhances the production of a species $Y$. $Y$ in turn, which gets rapidly degraded, stabilizes the activated form $X_p$, thus resulting in positive feedback loop (Fig. 3A).

The resulting model consists of three different cellular structures, membrane, cytoplasm, nucleus, which have to be taken into account plus the extracellular space.

We implemented two different models. In both models, the compartments describing the nucleus and cytoplasm were coupled using

$$X_{P,nuc} = X_{P,Cyt}$$

$$Y_{nuc} = Y_{Cyt}$$

as BCs. However, the models differed in the implementation of the membrane, which was either implemented as separate compartment or as boundary as described previously. If modeled as boundary, flux BCs were used to model the consumption of ligand $L$ for the formation of complex $C$, and for the consumption of species $X$ in the formation of its activated form $X_p$.

The models were simulated without or with the positive feedback from $Y$ on $X_p$. As observed for the simpler 2D model, the time courses of the species, if integrated over their respective domains, do not show differences (without feedback Fig. 3 B,D; with feedback Fig. 3 C,E). In both models, the feedback results, as expected, in higher levels for species $X_p$ and $Y$.

However, differences can be observed if the concentrations of the species of the separate compartments are considered. Due to the comparable big volume of the membrane, if modeled explicitly, compared to the volume of the cytoplasm, and the constraint that the concentration should be the same on both sides of the boundary, a considerable amount of $X_p$ can now be found in the membrane at all times (Fig. 3F). Furthermore, the solution of the model in which the membrane is explicitly modeled is computationally much more costly due to the thin layer describing the membrane.

## 5. Outlook

Within this manuscript we compared different implementations of compartmentalization and boundary reactions in COMSOL Multiphysics. We conclude that for the implementation of different compartments, an approach should be taken in which the PDEs are defined on each domain separately and coupled subsequently rather than an implementation using global PDEs.

In contrast to that, reactions that happen at a surface or in a, compared to the other compartments, very small compartment, should be implemented using the *boundary PDE*s interface and coupled to the neighboring compartments using flux BCs. Using this approach it would be also possible to implement species which are first bound to a membrane, but after e.g. activation are able to diffuse into the neighboring compartment.

Here, we have used these approaches only for the modeling of single cells. In principle, the implementation of multiple cells in a tissue amounts to nothing more than the coupling of more compartments as previously implemented in a model of lung branching morphogenesis [7]. The main limitation, however, is computational efficiency. More detailed modeling of tissue will therefore require the further optimization of solver options to implement multiple interacting cells without a dramatic increase in computational time.

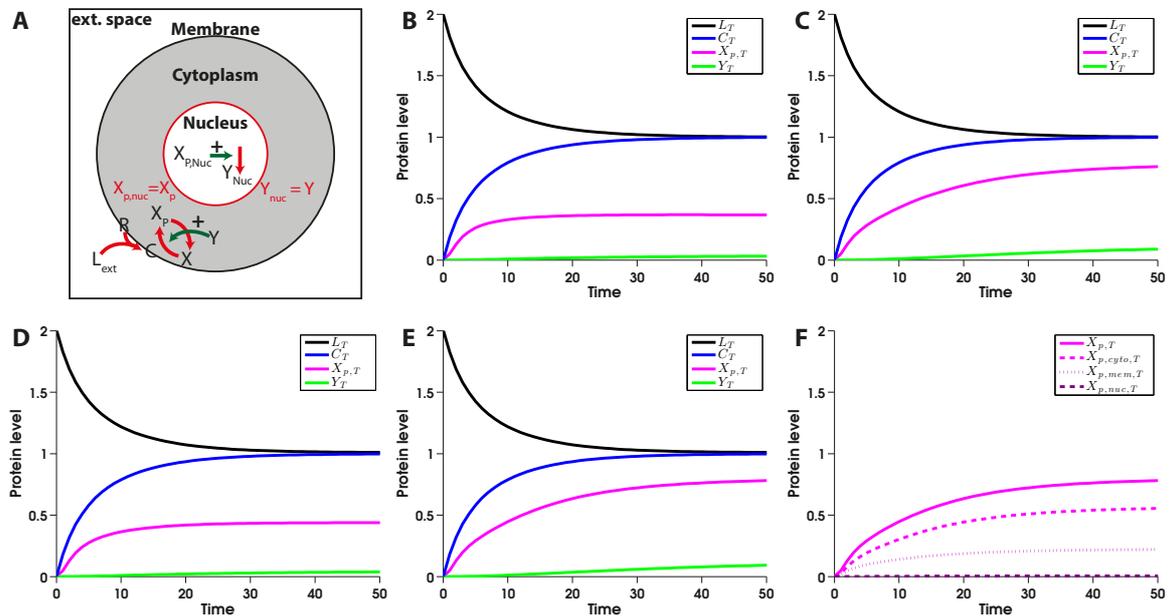

**Figure 3. A cell-based signaling model.** A) Scheme of the reactions present in the model, the BCs, the localization of the different species and their interactions. The scheme is shown as a cut plane through the 3D model geometry. B,C) Time courses for implementation of membrane as a boundary and use of flux BCs without (B) and with (C) positive feedback from $Y$ on $X_p$. D-F) Membrane implemented as separate volume and species coupled using constraints as BCs. Time courses show concentration of species over time without (D) and with (E) positive feedback from $Y$ on $X_p$ and the contribution of the different subspecies of $X_p$ on its total level (F). All protein levels are given as sum of all subspecies integrated over their respective compartments.

Furthermore, so far, we are not able to model cell division using COMSOL Multiphysics. An increase in tissue size would therefore need to be based on cell volume expansion only.

## 7. Acknowledgements

The authors acknowledge funding from the SNF Sinergia grant "Developmental engineering of endochondral ossification from mesenchymal stem cells", a SystemsX RTD on Forebrain Development, and a SIB PhD Fellowship to J.V.

## 8. Appendix

### 8.1 Model equations for cell-based signaling model

#### 8.1.1 Membrane implemented as boundary

*If not stated otherwise, no flux BCs were applied.*
(1) extracellular space

$$\frac{\partial L}{\partial t} = D_L \nabla^2 L, \quad BC: \frac{\partial L}{\partial n}|_{Mem} = -k_{on} \cdot R \cdot L$$

(2) Membrane

$$\frac{\partial R}{\partial t} = D_R \nabla^2 R - k_{on} \cdot R \cdot L, \quad \frac{\partial C}{\partial t} = D_C \nabla^2 C + k_{on} \cdot R \cdot L$$

(3) Cytoplasm

$$\frac{\partial X}{\partial t} = D_X \nabla^2 X + k_{dp} \cdot X_p \cdot \frac{K_i^n}{K_i^n + Y}, \quad BC: \frac{\partial X}{\partial n}|_{Mem} = -k_{act} \cdot C \cdot X$$

$$\frac{\partial X_p}{\partial t} = D_X \nabla^2 X_p - k_{dp} \cdot X_p \cdot \frac{K_i^n}{K_i^n + Y}, \quad BC: \frac{\partial X_p}{\partial n}|_{Mem} = +k_{act} \cdot C \cdot X$$

$$\frac{\partial Y}{\partial t} = D_X \nabla^2 Y - k_{deg} \cdot Y$$

(4) Nucleus

$$\frac{\partial X_{pn}}{\partial t} = D_X \nabla^2 X_{pn} - k_{dp} \cdot X_{pn} \cdot \frac{K_i^n}{K_i^n + Y},$$

$$BC: X_p |_{NucMem} = X_{pn} |_{NucMem}$$

$$\frac{\partial Y_n}{\partial t} = D_X \nabla^2 Y_n + k_{trans} \cdot X_{pn} - k_{deg} \cdot Y_n,$$

$$BC: Y_n |_{NucMem} = Y |_{NucMem}$$

#### 8.1.2 Membrane implemented as volume

If membrane was implemented as volume (thickness 1/10 of the sphere representing cytoplasm) $L_{mem}$, $X_{p,mem}$, $X_{mem}$ and $Y_{mem}$ were included explicitly in the model and coupled using constraints on the boundary (No flux BC on all boundaries).

#### 8.1.3 Without feedback

To simultate the model without feedback from *Y* on *X*, the term $\frac{K_i^n}{K_i^n + Y}$ was excluded from all equations.

#### 8.1.4 Parameters

$D_L=100$, $D_R = 1$, $D_X = 0.01$, $k_{on} = 0.1$, $k_{dp} = 1$, $k_{act} = 0.05$, $k_{trans} = 1$, $k_{deg} = 0.05$, $K_i = 0.5$, $n = 2$;